\documentclass[runningheads]{llncs}
\usepackage[table]{xcolor}
\usepackage{ragged2e}
\usepackage{uppaal}
\usepackage[labelformat=simple]{subcaption}

\usepackage{cite}
\usepackage{interval}
\usepackage{svg}
\usepackage{tikz}
\usepackage[inline]{enumitem}
\usepackage{wrapfig}
\usetikzlibrary{positioning, fit, shapes.geometric, arrows}
\usepackage[T1]{fontenc}
\usepackage{graphicx}

\usepackage{%
    algorithm,
    amsmath,
    amssymb,
    amsthm,
    bbm,
    subcaption,
    xspace,
    booktabs
}
\usepackage{todonotes}

\usepackage [noend]      {algpseudocode}
  \algrenewcommand{\algorithmicrequire}{Input}
  \algrenewcommand{\algorithmicensure}{Output}
\usepackage {hyperref}
\usepackage [capitalize] {cleveref} 

\usepackage{color}

\tikzset {%
    arrow/.style = {%
        thick,
        ->,
        >=stealth
    }
}

\graphicspath{{./imgs/}}

\newcommand{\code}[1] {\textcolor{wisteria}{\tt{#1}}}

\newcommand \RR    {\ensuremath {\mathbb R}\xspace}

\newcommand \NN {\ensuremath{\mathbb{N}}\xspace}

\newcommand \Tcal  {\ensuremath {\mathcal T}\xspace}

\newcommand \caap          {\textsc {Caap}\xspace}

\newcommand \tiga          {\textsc {Uppaal Tiga}\xspace}
\newcommand \ourtool       {\textsc{Uppaal Coshy}\xspace}
\newcommand \tempest       {\textsc{Tempest}\xspace}
\newcommand \prism         {\textsc{Prism}\xspace}
\newcommand{\mdp}{\ensuremath{\mathcal{M}}\xspace}
\newcommand{\states}{\ensuremath{S}\xspace}
\newcommand{\actions}{\ensuremath{A}\xspace}
\newcommand{\transitionfunction}{\ensuremath{T}\xspace}
\newcommand{\run}{\ensuremath{\pi}\xspace}
\newcommand{\strategy}{\ensuremath{\sigma}\xspace}
\newcommand{\stochasticstrategy}{\ensuremath{\bar{\sigma}}\xspace}
\newcommand{\strategyT}{\ensuremath{\nu}\xspace}
\newcommand{\prop}{\ensuremath{\varphi}\xspace}
\newcommand{\granularity}{\ensuremath{\gamma}\xspace}
\newcommand{\offset}{\ensuremath{\omega}\xspace}
\newcommand{\partit}{\ensuremath{\mathcal{P}}\xspace}
\newcommand{\grid}{\ensuremath{\partit_{\granularity}^{\offset}}\xspace}
\newcommand{\cell}{\ensuremath{C}\xspace}
\newcommand{\region}{\ensuremath{R}\xspace}
\newcommand{\cout}{\ensuremath{\cell_\textit{out}}\xspace}
\newcommand{\transitionsystem}{\ensuremath{\mathcal{T}_{\mdp,\granularity,\offset}}\xspace}
\newcommand{\safecellsinit}{\ensuremath{\mathcal{C}_\prop^0}\xspace}
\newcommand{\safecells}{\ensuremath{\mathcal{C}_\prop}\xspace}
\newcommand{\cc}[1]{\multicolumn{1}{c}{\textbf{#1}}}

\definecolor{turquoise}{HTML}{1ABC9C}
\definecolor{emerald}{HTML}{2ECC71}
\definecolor{peterriver}{HTML}{3498DB}
\definecolor{amethyst}{HTML}{9B59B6}
\definecolor{wetasphalt}{HTML}{34495E}

\definecolor{greensea}{HTML}{16A085}
\definecolor{nephritis}{HTML}{27AE60}
\definecolor{belizehole}{HTML}{2980B9}
\definecolor{wisteria}{HTML}{8E44AD}
\definecolor{midnightblue}{HTML}{2C3E50}

\definecolor{sunflower}{HTML}{F1C40F}
\definecolor{carrot}{HTML}{E67E22}
\definecolor{alizarin}{HTML}{E74C3C}
\definecolor{clouds}{HTML}{ECF0F1}
\definecolor{concrete}{HTML}{95A5A6}

\definecolor{orange}{HTML}{F39C12}
\definecolor{pumpkin}{HTML}{D35400}
\definecolor{pomegranate}{HTML}{C0392B}
\definecolor{silver}{HTML}{BDC3C7}
\definecolor{asbestos}{HTML}{7F8C8D}

\begin{document}
\title{\ourtool: Automatic Synthesis of \\ Compact Shields for Hybrid Systems}
\titlerunning{Automatic Synthesis of Compact Shields for Hybrid Systems}
%
\author{Asger~Horn~Brorholt\inst{1} \and
Andreas~Holck~H{\o}eg-Petersen\inst{1} \and
Peter~Gj{\o}l~Jensen\inst{1} \and
Kim~Guldstrand~Larsen\inst{1} \and
Marius~Miku\v{c}ionis\inst{1} \and
Christian~Schilling\inst{1} \and
Andrzej~W\k{a}sowski\inst{2}}
%
%
\authorrunning{Brorholt, H{\o}eg-Petersen, Jensen, Larsen, Miku\v{c}ionis, Schilling, W\k{a}sowski}
%
\institute{Aalborg University, 9220 Aalborg, Denmark \\
\email{\{asgerhb,ahhp,pgj,kgl,marius,christianms\}@cs.aau.dk}
\and
IT University of Copenhagen, 2300 Copenhagen, Denmark \\
\email{wasowski@itu.dk}}
\maketitle              
\begin{abstract}
We present \ourtool, a tool for automatic synthesis of a safety strategy---or \emph{shield}---for Markov decision processes over continuous state spaces and complex hybrid dynamics. The general methodology is to partition the state space and then solve a two-player safety game~\cite{DBLP:conf/vecos/BrorholtJLLS23}, which entails a number of algorithmically hard problems such as reachability for hybrid systems. The general philosophy of \ourtool is to approximate hard-to-obtain solutions using simulations. Our implementation is fully automatic and supports the expressive formalism of \uppaal models, which encompass stochastic hybrid automata.

The precision of our partition-based approach benefits from using finer grids, which however are not efficient to store. We include an algorithm called \caap to efficiently compute a compact representation of a shield in the form of a decision tree, which yields significant reductions.

\keywords{Shield synthesis \and \uppaal\ \and Decision tree.}
\end{abstract}

\section{Introduction}

%
%
%

In prior work, we proposed an algorithm to synthesize \emph{shields} (i.e., nondeterministic safety strategies) for Markov decision processes with hybrid dynamics~\cite{DBLP:conf/vecos/BrorholtJLLS23}.
The algorithm partitions the state space into finitely many cells and then solves a two-player safety game, where it uses approximation through simulation to efficiently tackle algorithmically hard problems.
In this tool paper, we present our implementation \ourtool, which is fully integrated in \uppaal, offering an automatic tool%
\footnote{Available at \url{https://uppaal.org/features/\#coshy}}
that supports the expressive \uppaal modeling formalism, including reinforcement learning under a shield.

Our algorithm represents a shield by storing the allowed actions for each cell individually, which results in a large data structure.
Since many neighboring cells allow the same actions in practice, as a second contribution, we propose a new algorithm called \caap to compute a compact representation in the form of a decision tree.
We demonstrate that this algorithm leads to significant reductions as part of the workflow in \ourtool.

\subsection{Related Tools for Shield Synthesis and Compact Representation}

\paragraph{Shielding.}
Shields are obtained by solving games, for which there exist a wide selection of tools for discrete state spaces~\cite{DBLP:conf/cav/ChatterjeeHJR10, DBLP:conf/tacas/ChatterjeeHJS11, DBLP:conf/cav/KwiatkowskaN0S20}.
%
Notably, \href{https://tempest-synthesis.org/}{\tempest}~\cite{DBLP:conf/atva/PrangerKPB21} synthesizes shields for discrete systems and facilitates learning through integration with \prism~\cite{DBLP:conf/cav/KwiatkowskaNP11}.
%
\tiga synthesizes shields for timed games~\cite{DBLP:conf/cav/BehrmannCDFLL07}.

In contrast, our tool applies to a richer class of models, including stochastic hybrid systems with non-periodic control and calls to external C libraries.

One benefit of our tool is the full integration with \stratego~\cite{DBLP:conf/tacas/DavidJLMT15} to directly use the synthesized shield in reinforcement learning.

\medskip

\paragraph{Decision trees.}
Encoding strategies as decision trees is a popular approach to achieving compactness and interpretability~\cite{DuLH20,Quinlan96,BreimanFOS84,DBLP:conf/hybrid/AshokJJKWZ20,DBLP:conf/tacas/AshokJKWWY21}.
However, these works focus on creating approximate representations from tabular data.
For a fixed set of predicates, the smallest possible tree can be obtained by enumeration techniques~\cite{DBLP:journals/jmlr/DemirovicLHCBLR22,DemirovicSL25}.
In contrast, our method transforms a given decision tree into an \emph{equivalent} decision tree.
Our method is specifically designed to efficiently cope with strategies of many axis-aligned decision boundaries.

\section{Shield Synthesis for Hybrid Systems}
\label{sect:shielding_algorithm}

In this section, we recall a general shield synthesis algorithm for hybrid systems outlined in prior work~\cite{DBLP:conf/vecos/BrorholtJLLS23}.
We start by recalling the formalism for control systems.

\subsection{Euclidean Markov Decision Processes}

\begin{definition}[Euclidean Markov decision process~\cite{DBLP:conf/atva/JaegerJLLST19}]\label{def:emdp}
A $k$-dimensional \emph{Euclidean Markov decision process} (EMDP) is a tuple $\mdp=(\states, \actions, \transitionfunction)$ where
\begin{itemize}
    \item $\states \subseteq \RR^k$ is a closed and bounded subset of the $k$-dimensional Euclidean space,
    \item $\actions$ is a finite set of actions, and
    \item $\transitionfunction \colon \states \times \actions \rightarrow (\states \rightarrow \RR_{\geq0} )$ maps each state-action pair $(s,a)$ to a probability density function over $\states$, i.e., we have $\int_{s'\in \states} \transitionfunction(s,a)(s') ds'=1$.
\end{itemize}
\end{definition}

For simplicity, the state space~$\states$ is continuous. However, the extension to discrete variables, e.g., locations of hybrid components, is straightforward.
Since optimizing strategies is not our focus, we do not formally introduce the notion of cost and rely on the reader's intuition.
(See~\cite{DBLP:conf/vecos/BrorholtJLLS23} for details.)

A \emph{run}~$\run$ of an EMDP is an alternating sequence $\run = s_0a_0s_1a_1\dots$ of states and actions such that $T(s_i,a_i)(s_{i+1})>0$ for all $i\geq0$.
A (memoryless) stochastic \emph{strategy} for an EMDP is a function $\stochasticstrategy \colon\states\rightarrow(\actions\rightarrow[0,1])$, mapping a state to a probability distribution over the actions.
A run $\run = s_0a_0s_1a_1\dots$ is an \emph{outcome} of $\stochasticstrategy$ if $\stochasticstrategy(s_i)(a_i) > 0$  for all $i \ge 0$.
Similarly, a (memoryless) nondeterministic strategy is a function $\strategy \colon\states\rightarrow2^{\actions}$, mapping a state to a set of actions.
A run $\run = s_0a_0s_1a_1\dots$ is an outcome of $\strategy$ if $a_i\in \strategy(s_i)$  for all $i \ge 0$.

\medskip

A \emph{safety property} (or invariant) $\prop$ is a set of states $\prop\subseteq\states$.
A run $\run = s_0a_0s_1a_1\dots$ is \emph{safe} with respect to $\prop$ if $s_i\in\prop$ for all $i\geq 0$.
A nondeterministic strategy~$\strategy$ is a \emph{shield} with respect to~$\prop$ if all outcomes of~$\strategy$ are safe.

\subsection{Running Example (Bouncing Ball)}

\begin{figure}[tb]
    \centering
    \begin{subfigure}{0.31\linewidth}
        \centering
        \includegraphics[width=\linewidth]{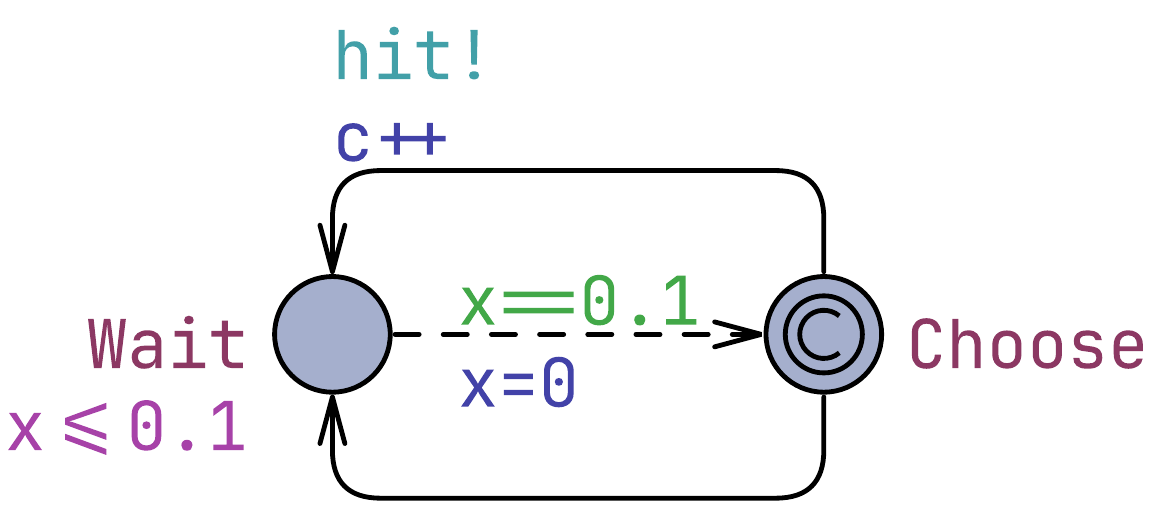}
        \caption{Player component.}
        \label{fig:player}
    \end{subfigure}
    \hfill
    \begin{subfigure}{0.65\linewidth}
        \centering
        \includegraphics[width=\linewidth]{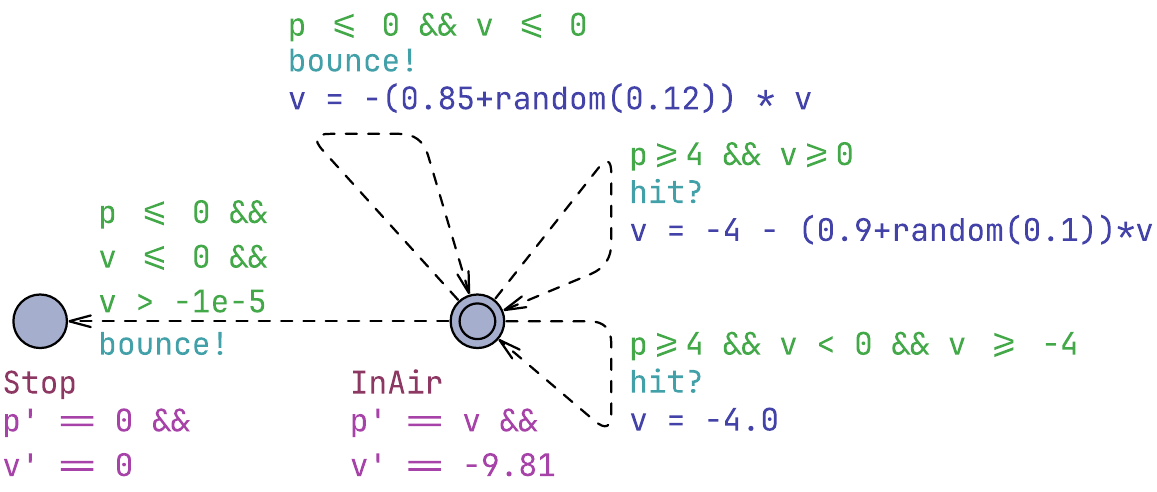}
        \caption{Ball component.}
        \label{fig:ball}
    \end{subfigure}
    \caption{The \emph{bouncing ball} modeled in \uppaal.}
    \label{fig:uppaalball}
\end{figure}

We introduce our running example: a \emph{bouncing ball} that can be hit by a player to keep it bouncing~\cite{DBLP:conf/vecos/BrorholtJLLS23,DBLP:conf/atva/JaegerJLLST19}.
We shortly explain our two-component \uppaal model.
The player component is shown in \cref{fig:player}.
In the (initial) location \uppLoc{Choose}, there are two available control actions (solid lines).
The player chooses every 0.1 seconds (enforced by the clock~\code{x}).
The \uppSync{hit} action (upper edge) attempts to hit the ball, and increments the cost counter~\code{c} to be used for reinforcement learning in \cref{sect:bb_queries}.
The other action (lower edge) does not attempt to hit the ball.

The ball component, shown in \cref{fig:ball}, is described by two state variables, position~\code{p} and velocity~\code{v}, which evolve according to the ordinary differential equations shown below the initial location \uppLoc{InAir}.
The two dashed edges on the right model a successful \uppSync{hit} action, which is only triggered if the ball is high enough (four meters or higher above the ground); they differ in whether the ball is currently jumping up or falling down.
The two dashed edges on the left model a \uppSync{bounce} on the ground.
The ball bounces back up with a random dampening (upper edge) or goes to the state \uppLoc{Stop} if the velocity is very low (lower edge).
In the following, we shall see how to obtain a shield that enforces the safety property that \uppLoc{Stop} is never reached, i.e., $\prop = \{s \mid \text{Ball is not in \uppLoc{Stop} in } s\}$.

\subsection{Partition-Based Shield Synthesis}\label{sect:partitioning}

Since an EMDP consists of infinitely many states, we employ a finite-state abstraction.
For that, we partition the state space $\states \subseteq \RR^k$ with a regular \emph{rectangular} grid. (In~\cite{DBLP:conf/vecos/BrorholtJLLS23}, we only allowed a grid of uniform size in all dimensions.)
Formally, given a (user-defined) granularity vector~$\granularity \in \RR^k$ and offset vector~$\offset \in \RR^k$, we partition the state space into disjoint \emph{cells} of equal size.
Each cell~$\cell$ is the Cartesian product of half-open intervals $\rinterval{\offset_i + p_i \granularity_i}{~\offset_i + (p_i + 1)\granularity_i}$ in each dimension~$i$, for cell index~$p \in \NN^k$.
We define the \emph{grid} as the set $\grid = \lbrace \cell \mid \cell \cap \states \neq \emptyset \rbrace$ of all cells that overlap with the bounded state space.
Note the number of cells will depend on $\granularity$.
For each $s\in\states$, $[s]_{\grid}$ denotes the unique cell containing~$s$.

An EMDP~$\mdp$, a granularity vector~$\granularity$ and offset vector $\offset$ induce a finite labeled transition system $\transitionsystem=(\grid,\actions,\rightarrow)$, where
\begin{equation}\label{eq:cell_reachability}
    \cell \xrightarrow[]{a}\cell' \iff \exists s\in\cell.\ \exists s'\in\cell'.\ \transitionfunction(s,a)(s')>0.
\end{equation}

Given a safety property $\prop \subseteq \states$ and a grid~$\grid$, let $\safecellsinit = \{\cell \in{\grid} \mid \cell \subseteq \prop\}$ denote those cells that are safe in zero steps.
We define the set of \emph{safe cells} as the maximal set~$\safecells$ such that
\begin{equation}\label{eq:safecells}
	\safecells = \safecellsinit \cap \{ \cell \in \grid \mid \exists a \in \actions.\ \forall \cell' \in \grid.\ \cell\xrightarrow{a}\cell' \implies \cell'\in \safecells \}.
\end{equation}

Given the finiteness of $\grid$ and monotonicity of \cref{eq:safecells}, $\safecells$ may be obtained in a finite number of iterations using Tarski's fixed-point theorem~\cite{Tarski55}.

A (nondeterministic) strategy for $\transitionsystem$ is a function $\strategyT:\grid\rightarrow 2^{\actions}$.
The most permissive shield~$\strategyT_\prop$ (i.e., safe strategy) obtained from $\safecells$~\cite{BernetJW02} is given by
\[
	\strategyT_\prop(\cell)=\{a \in \actions \mid \forall\cell' \in \grid.\ \cell\xrightarrow{a}\cell'\implies \cell'\in\safecells\}.
\]

A shield~$\strategyT$ for~$\transitionsystem$ induces a shield~$\strategy$ for~$\mdp$ in the standard way~\cite{DBLP:conf/vecos/BrorholtJLLS23}:

\begin{theorem}\label{thm:safety_transfer}
Given an EMDP~$\mdp$, a safety property~$\prop \subseteq \states$, and a grid~$\grid$, if~$\strategyT$ is a shield for $\transitionsystem$, then $\strategy(s) = \strategyT([s]_{\grid})$ is a shield for~$\mdp$.
\end{theorem}

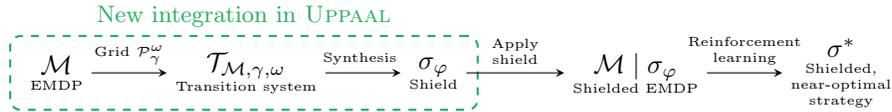
\begin{figure}[tb]
    \centering
    \begin{tikzpicture}[
    node distance=1cm, 
    every node/.style={align=center},
    >=stealth,
    font=\tiny
]
    \node[] (EMDP) {\normalsize $\mathcal{M}$ \\  EMDP};
    \node[right=of EMDP] (TAM) {\normalsize $\transitionsystem$ \\  Transition system};
    \node[right=of TAM] (sigma_phi) {\normalsize $\sigma_{\varphi}$ \\ Shield};
    \node[right=13mm of sigma_phi] (M_shielded) {\normalsize $\mathcal{M} \mid \sigma_{\varphi}$ \\ Shielded EMDP};
    \node[right=of M_shielded] (sigma_star) {\normalsize $\sigma^*$ \\ Shielded, \\ near-optimal \\ strategy};
    
    \node[draw=nephritis, dashed, thick, inner sep=1.5mm, fit=(EMDP) (sigma_phi), label={[nephritis]above: {\small New integration in \uppaal}}, rounded corners] {};

    \draw[->] (EMDP) -- (TAM) node[midway, above] { Grid $\grid$};
    \draw[->] (TAM) -- (sigma_phi) node[midway, above] { Synthesis };
    \draw[->] (sigma_phi) -- (M_shielded) node[midway, above] { Apply \\ shield };
    \draw[->] (M_shielded) -- (sigma_star) node[midway, above] { Reinforcement \\ learning};
    
\end{tikzpicture}
    \caption{Workflow for obtaining a near-optimal shielded strategy in \uppaal.}
    \label{fig:workflow}
\end{figure}

\cref{fig:workflow} shows the overall workflow of the shield synthesis and how the shield can later be used to (reinforcement-) learn a near-optimal strategy \emph{under this shield}.
The green box marks the steps that we newly integrated in \uppaal.

\medskip

\begin{figure}[tb]
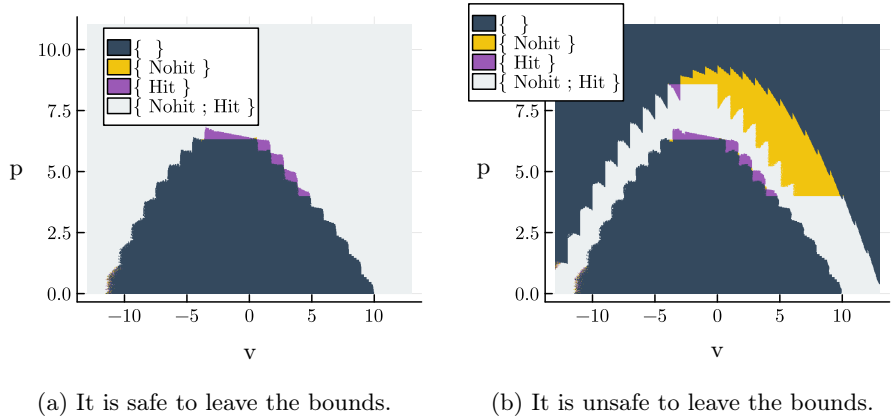

    \centering
    \begin{subfigure}[b]{0.49\textwidth}
        \centering
        \includesvg[width=\linewidth]{Graphics/BB.svg}
        \caption{It is safe to leave the bounds.}
        \label{fig:leave_bounds}
    \end{subfigure}
    \hfill
    \begin{subfigure}[b]{0.49\textwidth}
        \centering
        \includesvg[width=\linewidth]{Graphics/BB_constrained_to_grid.svg}
        \caption{It is unsafe to leave the bounds.}
        \label{fig:stay_in_bounds}
    \end{subfigure}
    \caption{Two shields for the \emph{bouncing ball}. Colors represent the allowed actions in the corresponding state of velocity~$v$ and position~$p$ while in location~\code{InAir}.}
    \label{fig:bbshields}
\end{figure}

For the \emph{bouncing ball}, we will obtain the shield shown in \cref{fig:leave_bounds}.
To effectively implement the aforementioned approach, there are additional challenges which we address in the following section.

\section{Effective Implementation of Shield Synthesis}
\label{sect:shielding_implementation}

In this section, we discuss our implementation of the approach to synthesize a shield as outlined in \cref{sect:shielding_algorithm} in \ourtool.
In particular, a practical implementation faces the following two main challenges.

First, we receive the safety property~$\prop$ in the form of a user query (described in \cref{appendix:query_syntax}).
Thus, the definition of the cells~$\safecellsinit$ that are immediately safe generally requires symbolic reasoning, which is not readily available. Instead, we check a finite number of states within each cell, which we describe in \cref{sect:initial_safe}.

Second, determining \cref{eq:cell_reachability} requires to solve reachability questions for infinitely many states.
While this can be done for simple classes of systems, we deal with very general systems (e.g., nonlinear hybrid dynamics), for which reachability is undecidable~\cite{DFPP18}.
This motivated us to instead compute an approximate solution, which we outline in \cref{sect:reachability}.

Thanks to the above design decisions, our implementation is fully automatic and supports the expressive formalism of general \uppaal models (e.g., stochastic hybrid automata with calls to general C code).

We also identified further practical challenges, which we address in the later parts of this section.
\cref{def:emdp} requires a bounded state space, but it is for instance difficult to determine upper bounds for the position and velocity of the \emph{bouncing ball}; in \cref{sect:unbounded}, we explain how we treat such cases in practice.
In \cref{sect:missing_variables}, we discuss an optimization to omit redundant dimensions.

\subsection{Determining Initial Safe Cells}\label{sect:initial_safe}

We apply \emph{systematic sampling} from a cell, i.e., samples are not drawn at random.
Rather, we uniformly cover the cell with $n^k$ samples,
where $n \in \NN, n \neq 0$ is a user-defined parameter.
Recall from \cref{sect:partitioning} that a cell~$\cell$ of a grid~$\grid$ is rectangular and defined by an index vector~$p$, an offset~$\offset$ and a granularity vector~$\gamma$, all of dimension~$k$.
Let $\delta_i = \frac{\granularity_i}{n - 1}$ be the distance between two samples in dimension~$i$ when $n > 1$, and $\delta_i = 0$ otherwise.
For any cell, we define the corresponding set of samples as $\left\{
(\offset_1 + p_1 \gamma_1 + q_1 \delta_1, \dots, \offset_k + p_k \gamma_k + q_k \delta_k) \mid q_i \in \{ 0, 1, \dots, n - 1 \}
\right\}$.
To account for the open upper bounds, we subtract a small number $\epsilon > 0$ from the highest samples.
%
%
An example of a two-dimensional set of samples for $n=4$ is shown as the dark blue points inside the light blue cells in \cref{fig:reachability}.

We note that the above only applies to continuous variables.
Our implementation treats discrete variables (e.g., component locations) in the natural way.

\medskip

Finally, to approximate the set~$\safecellsinit$, we draw samples from each cell and check for each sample whether it violates the specification.
A cell is added to~$\safecellsinit$ only if all samples in that cell satisfy the specification.

For the \emph{bouncing ball}, the ball should never be in the \uppLoc{Stop} location.
Since the location is a discrete variable, and each cell only belongs to one location, checking a single sample from a cell~$\cell$ already determines whether $\cell \in \safecellsinit$.
Thus, our approach is exact and efficient in the common case where the safety property is given via an error location.

\subsection{Determining Reachability}\label{sect:reachability}

We approximate cell reachability $\cell \xrightarrow{a} \cell'$, as defined in \cref{eq:cell_reachability}, similarly to~\cite{DBLP:conf/vecos/BrorholtJLLS23} but adapted to work in \uppaal.
In a \uppaal model, actions $a \in \actions$ correspond to controllable edges (indicating that the controller can act).

For each cell~$\cell$ and action~$a \in \actions$, we iterate over all sampled states~$s$ (as described before) and select the edge corresponding to $a$, which gives us a new state~$s'$; starting from $s'$, we simulate the environment (using the built-in simulator in \uppaal) until a state $s''$ is reached in which the controller has the next choice (i.e., multiple action edges are enabled) again.\footnote{Where~\cite{DBLP:conf/vecos/BrorholtJLLS23} required a fixed control period, \ourtool supports non-periodic control. We include an example of this feature in \cref{appendix:nonperiodic}.}
Thus, $s''$ is a witness to add the corresponding cell~$[s'']_{\grid}$ to the transition relation $\cell \xrightarrow{a} [s'']_{\grid}$.
Assuming the simulator is numerically sound, the resulting transition system underapproximates~$\transitionsystem$.
As observed in~\cite{DBLP:conf/vecos/BrorholtJLLS23}, the more simulations are run, the more likely do we obtain the true solution.
To check whether this underapproximation is sufficiently accurate, the existing queries for statistical model checking in \uppaal can be used, as we shall see in \cref{sect:evaluation}.

\medskip

In general, two simulations starting in the state~$s$ may not yield the same state~$s''$ due to stochasticity.
In~\cite{DBLP:conf/vecos/BrorholtJLLS23}, stochasticity was treated as additional dimensions over which to sample (systematically).
This was possible by manually crafting the reachability sampling for each model.
Detecting stochastic behavior in \uppaal models automatically turned out to be difficult due to the rich formalism.
Instead, we decided to simply let the simulator sample from the stochastic distribution.
As a side effect, this new design allows us to support stochasticity with general distributions, particularly with unbounded support.

Since this design may generally miss some corner-case behavior, we expose a user-defined parameter~$m$ to control the number of times sampling is repeated.

\begin{figure}[tb]
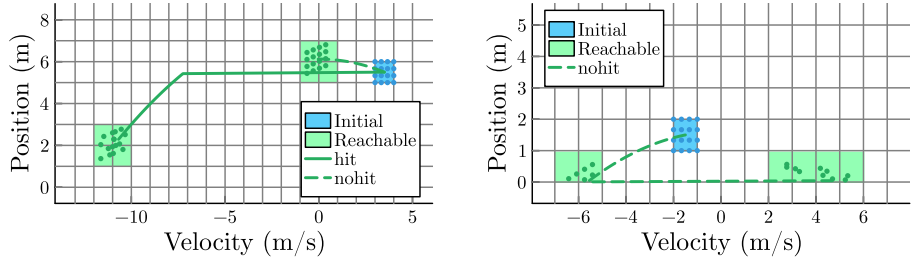

    \begin{subfigure}{0.48\linewidth}
      \centering
      \includesvg[width=\linewidth]{"Graphics/Reachability1.svg"}
      \caption{The ball is rising and high enough to be hit. When the ball is hit, the outcome is partially random.}
      \label{fig:reachabilityA}
    \end{subfigure}
    \hfill
    \begin{subfigure}{0.48\linewidth}
      \centering
      \includesvg[width=\linewidth]{"Graphics/Reachability2.svg"}
      \caption{The ball is too low to be hit, but it bounces off the ground. The velocity loss upon a bounce is partially random.}
      \label{fig:reachabilityB}
    \end{subfigure}
    \caption{Example of a grid for the \emph{bouncing ball}.
    By sampling from the initial cell (blue) and simulating the dynamics, we discover reachable cells (green).}
    \label{fig:reachability}
\end{figure}

We illustrate the reachability approximation for the \emph{bouncing ball} in \cref{fig:reachability} for $n=4$ (number of samples per dimension) and $m=1$ (number of re-sampling). When the ball moves through the air, it behaves deterministically. In \cref{fig:reachabilityA}, when the ball is not hit, we obtain successor states that keep a regular ``formation'' (top right green dots).
When the ball is hit, the successor states are affected by randomness (bottom left green dots).
\cref{fig:reachabilityB} shows a similar randomized effect when the ball touches the ground.

\subsection{Generalization to Unbounded State Spaces}
\label{sect:unbounded}

\cref{def:emdp} requires the state space to be bounded, but bounds can be hard to determine for some systems.
This includes the \emph{bouncing ball}, for which upper bounds for position and velocity are not immediately clear.
Indeed, if we consider the bounded state space where $p \in \interval{0}{11}$ and $v \in \interval{-13}{13}$, the system dynamics do not guarantee that the ball stays within these bounds.
If we plot velocity against position, as in \cref{fig:bbshields}, then a falling ball near the left end of the plot may leave the bounds on the left (because it becomes too fast).

Conceptually, our implementation deals with out-of-bounds situations by modifying the transition system.
All samples leading to a state outside the specified bounds go to a dummy cell $\cout$, for which all transitions lead back to itself.
A user-defined option with the following choices determines the behavior:

\begin{enumerate}
    \item Raise an error when reaching $\cout$ during simulation (default behavior).
    \item \label{it:always_safe} Include $\cout$ in $\cell_\prop^0$, i.e., leaving the bounds is always safe.
    \item \label{it:always_unsafe} Exclude $\cout$ from $\cell_\prop^0$, i.e., leaving the bounds is always unsafe.
    \item \label{it:auto} Automatically choose between options~\ref{it:always_safe} and~\ref{it:always_unsafe} using sampling.
\end{enumerate}

With Option~\ref{it:auto}, samples are taken outside the specified bounds, similar to \cref{sect:initial_safe}.
For the \emph{bouncing ball}, our tool samples states such as $(v=26$, $p=22$, $\code{Ball.Stop})$, even though these states may not be reachable in practice.
If any sample state is found to be unsafe, $\cout$ is considered unsafe, and safe otherwise.
The result of synthesizing a shield with this option is shown in \cref{fig:stay_in_bounds}.
In particular, that shield forbids to hit the ball when it is too fast, which ensures that it does not leave the bounds.
Alternatively, we obtain a more permissive shield by choosing Option~\ref{it:always_safe}, as shown in \cref{fig:leave_bounds} (and also \cref{fig:rand_period_shield}).

\subsection{Omitting Variables from Consideration}
\label{sect:missing_variables}

As emphasized in~\cite{DBLP:conf/aaai/AlshiekhBEKNT18}, a shield can be obtained from an abstract model that only simulates behaviors relevant to the safety specification.
For example, cost variables may only be relevant during learning.
While every variable in a model can be included in the partitioning, this is computationally demanding.

Therefore, we allow that variables are omitted from the grid specification.
However, this raises a new challenge when sampling a state from a cell, since a concrete state requires a value for each variable.
To address that, we set each omitted variable to the unique value of the initial state, which must always be specified in a \uppaal model.
Hence, the user must define the initial state such that the values of omitted variables are sensible defaults.
(Note that the initial state is ignored by the shield synthesis in all other respects.)

The choice not to include a variable in the grid must be made carefully, as this can change the behavior of the transition system and potentially lead to an unsound shield.
As a rule of thumb, it is appropriate to omit variables if they always have the same value when actions are taken, or if they are only relevant for keeping track of a performance value such as cost.

For the \emph{bouncing ball}, the player (\cref{fig:player}) is always in the location \uppLoc{Choose} when taking an action.
By setting \uppLoc{Choose} as the initial location, this component's location is not relevant to keep track of in the partitioning.
Moreover, the variable~\code{c} is used to keep track of cost and does not matter to safety.
Lastly, the clock variable~\code{x} is used to measure time until the next player action. It is always~$0$ when it is time for the player to act, and so it can also be omitted.

%
%

\section{Obtaining a Compact Shield Representation}%
\label{sec:maxParts}

In this section, we present a new technique for obtaining a compact representation of shields that stem from an axis-aligned state-space partitioning (as described in \cref{sect:partitioning}).
Here, we choose to represent the shield as a decision tree.
We note that we aim for a functionality-preserving representation, i.e., we transform a grid-based shield to an equivalent decision-tree-based shield.

Recall that each cell prescribes a set of allowed actions.
Let two cells be \emph{similar} if the shield assigns the same set of actions to them.
Our goal is to form (hyper)rectangular clusters of similar cells, which we call \emph{regions}; in other words, we aim to find a coarser partitioning.
In a nutshell, our approach works as follows.
Initially, we start from the finest partitioning where each cell is a separate region.
Then, we iteratively merge neighboring regions of similar cells, thereby obtaining a coarser partitioning, such that the resulting region is rectangular again.
We call our algorithm \caap (\textbf{C}oarsify \textbf{A}xis-\textbf{A}ligned \textbf{P}artitionings).

\subsection{Representation of Partitionings and Regions}

We start by noting that an axis-aligned partitioning of a state space $\states \subseteq \RR^k$ can be represented by a binary decision tree~$\Tcal$ where each leaf node is a set of actions and each inner node splits the state space with a predicate of the form $\rho(s) = s_i < c$, where~$s$ is a state vector, $s_i$ is a state dimension, and~$c \in \RR$.
Given a state~$s$, the tree evaluation, written $\Tcal(s)$, is defined as usual:
Start at the root node.
At an inner node, evaluate the predicate~$\rho(s)$.
If $\rho(s)=\top$, descend to the left child; otherwise, descend to the right child.
At a leaf node, return the corresponding set of actions.
We denote the partitioning induced by a decision tree $\Tcal$ as~$\partit _\Tcal$.
Our goal in this section is: given a decision tree \Tcal inducing a partitioning \( \partit _\Tcal \), find an equivalent but smaller decision tree.

\medskip

\begin{wrapfigure}{r}{0.16\textwidth}
\centering
\vspace{-3em}
\[
  \begin{tabular}{r@{\ }|@{\ }c@{\ }c@{\ }c@{\ }c}
    \text{} & 1 & 2 & 3 & 4 \\
    \midrule
    $s_1$ & 0 & 2 & 3 & 4 \\
    $s_2$ & 0 & 2 & 3 & 4
  \end{tabular}
\]
\vspace{-2em}
\label{fig:exampleMatrix}
\end{wrapfigure}

Given a tree~$\Tcal$, we store all bounds~$c$ of the predicates $s_i < c$ in a matrix~$M$ of~$k$ rows where the $i$-th row contains the bounds associated with state dimension~$s_i$ in ascending order. For example, consider the bounds in \cref{fig:expRulesOrg} and $M$ on the right.

We extract a bounds vector from~$M$ via an index vector $p \in \mathbb{N}^k$ such that the $i$-th entry of~$p$ contains the column index for the $i$-th row. In other words, the resulting vector consists of the values~$M_{i,p_i}$. For instance, $p = (1, 3)$ yields the vector $s^p = (0, 3)$ (row~$s_1$ column~$1$ and row~$s_2$ column~$3$). We can view this vector as a state in the state space given as $s^p = (M_{1,p_1}, \ldots, M_{k,p_k})$.

We define a region~$\region$ in terms of two index vectors $(p^{\min}, p^{\max})$ representing the minimal and maximal corner in each dimension. Then, increasing~$p^{\max}_i$ corresponds to expanding~$\region$ in dimension~$i$.

\subsection{Expansion of Rectangular Regions}

For an expansion to be legal, it must satisfy the following three
\textit{expansion rules}:

\begin{definition}[Expansion rules]\label{def:expansionRules}
    Let \( \region' \) be a candidate region for a new partitioning $\partit'$ derived
    from \( \partit_\Tcal \).
    Then $\region'$ is legal if it satisfies these three rules:

    \begin{enumerate}
        \item \label{it:rule1} All cells in region \( \region' \) have the same action set,

        \item \label{it:rule2} Region \( \region' \) does not intersect with other regions in \( \partit' \),

        \item \label{it:rule3} Region \( \region' \) does not cut any other region \( \region \) from the original partitioning \(\partit_\Tcal\) in two, i.e., the difference \( \region \setminus \region' \) is either empty or rectangular.
    \end{enumerate}
\end{definition}

\begin{figure*}[t!]
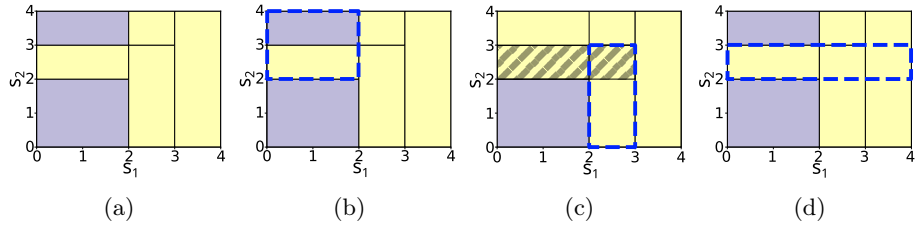

    \centering
    \begin{subfigure}{0.242\textwidth}
        \includesvg[width=\textwidth]{rules_org}%
        \caption{}%
        \label{fig:expRulesOrg}
    \end{subfigure}
    \begin{subfigure}{0.242\textwidth}
      \includesvg[width=\textwidth]{rule1}%
      \caption{}%
      \label{fig:expRules1}
    \end{subfigure}
    \begin{subfigure}{0.242\textwidth}
      \includesvg[width=\textwidth]{rule2}%
      \caption{}%
      \label{fig:expRules2}
    \end{subfigure}
    \begin{subfigure}{0.242\textwidth}
      \includesvg[width=\textwidth]{rule3}%
      \caption{}%
      \label{fig:expRules3}
    \end{subfigure}

  \caption{%
    Expansion example.
    Yellow and purple denote distinct actions.
    Striped regions have been fixed in previous iterations.
    The dashed border is the new candidate region~$\region'$.
    \subref{fig:expRulesOrg}~An input partitioning.
    \subref{fig:expRules1}~A violation of Rule~\ref{it:rule1}, since the expanded region contains different actions.
    \subref{fig:expRules2}~A violation of Rule~\ref{it:rule2}, since the expanded region overlaps with a striped area.
    \subref{fig:expRules3}~A violation of Rule~\ref{it:rule3}, since the expansion cuts the rightmost region into two new regions.
    }\label{fig:expansionRules}
\end{figure*}

\noindent
The first two cases are directly related to the definition of the problem, i.e., the produced partitioning should respect \Tcal and only have non-overlapping regions (see \cref{fig:expRules1} and \cref{fig:expRules2}).
The third case is required in order to ensure that in each iteration, the algorithm does not increase the overall number of regions when adding a region from the original partitioning to the new partitioning.
To appreciate this, consider the visualization in \cref{fig:expRules3}.
The candidate expansion cuts the rightmost region (given by \( (3,0) \) and \( (4,4) \)) in two such that the remainder would have to be represented by \textit{two} regions --- one given by \( ((3,0), (4,2)) \) and one given by $((3,3), (4,4))$.
Clearly, all three expansion rules of \cref{def:expansionRules} can be checked in time linear in the number of nodes of~$\partit_{\Tcal}$.

The rules induce a nondeterministic greedy algorithm for expanding regions.
This algorithm is included in \cref{appendix:caapalgorithm}.

\section{Case Studies and Evaluation}
\label{sect:evaluation}

In this section, we evaluate our implementation of \ourtool and \caap.
In \cref{sect:bb_queries}, we demonstrate a typical application.
In \cref{sect:benchmarks}, we benchmark the implementations on several models.

\subsection{A Complete Run of the Bouncing Ball}
\label{sect:bb_queries}

\begin{table}[tb]
    \rowcolors{2}{gray!25}{white}
    \caption{Queries run on the \emph{bouncing ball} model. All statistical results are given with a 99\% confidence interval.}
    \label{tab:bb_queries}
        \begin{tabular}{lll}
            \hline
            \textbf{\#} & \textbf{Query} & \textbf{Result} \\
            \hline

            1\hspace{10pt} & \begin{minipage}{0.78\linewidth} \begin{verbatim}
strategy efficient = minE(c) [<=120]
  {} -> {v, p} : <> time>=120
\end{verbatim} \end{minipage} &
            \checkmark \\

            2\hspace{10pt} & \begin{minipage}{0.78\linewidth} \begin{verbatim}
simulate [<=120]{ p, v } under efficient
\end{verbatim} \end{minipage} &
            \checkmark \\

            3\hspace{10pt} & \begin{minipage}{0.78\linewidth} \begin{verbatim}
E[<=120;100] (max: c) under efficient
\end{verbatim} \end{minipage} &
            $\approx 0$\\

            4\hspace{10pt} & \begin{minipage}{0.78\linewidth} \begin{verbatim}
Pr[<=120;10000] (<> Ball.Stop) under efficient
\end{verbatim} \end{minipage} &
            $\interval{0.9995}{1}$ \\

            5\hspace{10pt} & \begin{minipage}{0.78\linewidth} \begin{verbatim}
strategy shield = acontrol: A[] !Ball.Stop
  { v[-13, 13]:1300, p[0, 11]:550, Ball.location }
\end{verbatim} \end{minipage} &
            \checkmark \\

            6\hspace{10pt} & \begin{minipage}{0.78\linewidth} \begin{verbatim}
saveStrategy("/shield.json", shield)
\end{verbatim} \end{minipage} &
            \checkmark \\

            7\hspace{10pt} & \begin{minipage}{0.78\linewidth} \begin{verbatim}
strategy compact_shield = loadStrategy("/compact.json")
\end{verbatim} \end{minipage} &
            \checkmark \\

            8\hspace{10pt} & \begin{minipage}{0.78\linewidth} \begin{verbatim}
simulate [<=120]{ p, v } under compact_shield
\end{verbatim} \end{minipage} &
            \checkmark \\

            9\hspace{10pt} & \begin{minipage}{0.78\linewidth} \begin{verbatim}
strategy shielded_efficient = minE(c) [<=120]
  {} -> {v, p} : <> time>=120 under compact_shield
\end{verbatim} \end{minipage} &
            \checkmark\\

            10\hspace{10pt} & \begin{minipage}{0.78\linewidth} \begin{verbatim}
simulate [<=120]{ p, v } under shielded_efficient
\end{verbatim} \end{minipage} &
            \checkmark \\

            11\hspace{10pt} & \begin{minipage}{0.78\linewidth} \begin{verbatim}
E[<=120;100] (max: c) under shielded_efficient
\end{verbatim} \end{minipage} &
            $34.6 \pm 0.6$\\

            12\hspace{10pt} & \begin{minipage}{0.78\linewidth} \begin{verbatim}
Pr[<=120;10000] (<> Ball.Stop) under shielded_efficient
\end{verbatim} \end{minipage} &
            $\interval{0}{0.00053}$\\ 

        \end{tabular}
\end{table}

\cref{tab:bb_queries} shows a typical usage of \uppaal with a sequence of queries on the \emph{bouncing ball} example to produce a safe and efficient strategy (cf.\ \cref{fig:workflow}).
A detailed explanation of the new query syntax can be found in \cref{appendix:query_syntax}.
Documentation is also available online,%
\footnote{\url{https://docs.uppaal.org/language-reference/query-syntax/controller_synthesis/\#approximate-control-queries}}
including standard \uppaal queries.

\begin{figure}[b]
    \centering
    \begin{subfigure}[t]{0.32\textwidth}
        \centering
        \includegraphics[width=\linewidth]{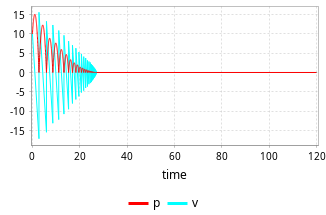}
        \caption{\code{efficient}}\label{fig:efficient}
    \end{subfigure}
    \hfill
    \begin{subfigure}[t]{0.32\textwidth}
        \centering
        \includegraphics[width=\linewidth]{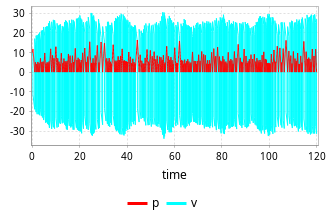}
        \caption{\code{compact\_shield}}\label{fig:safe}
    \end{subfigure}
    \hfill
    \begin{subfigure}[t]{0.32\textwidth}
        \centering
        \includegraphics[width=\linewidth]{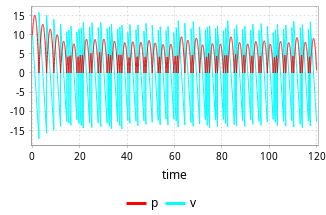}
        \caption{\code{shielded\_efficient}}\label{fig:shielded_efficient}
    \end{subfigure}
    \caption{\emph{Bouncing ball} simulations (position, velocity) under different strategies.}
    \label{fig:simulate}
\end{figure}

In Query~1, we train a strategy called \code{efficient}, which is only concerned with cost and does not consider safety.
Such a strategy is trivial: simply never pick the \uppSync{hit} action.
This is seen in Query~2, which simulates a single run of 120 seconds. It outputs position~\code{p} and velocity~\code{v}, which are visualized in \cref{fig:efficient}.
Query~3 statistically evaluates the strategy in 100 runs to estimate the expected value of~\code{c}.
The result ``$\approx 0$'' indicates that only this value was observed.
Query~4 estimates the probability of a run being unsafe to be in the interval $\interval{0.9995}{1}$ with 99\% confidence; in this case, as expected, all $10\,000$ runs were unsafe.

Query~5 synthesizes a shield \code{shield}.
The shield matches the one shown in \cref{fig:leave_bounds}.
In queries~6 and~7, the shield is converted to a compact representation by saving it to a file, calling the \caap implementation, and loading the result back into \uppaal.
The shield is simulated in Query~8, for which any of the allowed actions is selected randomly (this happens implicitly); while safe, this shielded but randomized strategy is not efficient and hits the ball more often than needed, as visualized in \cref{fig:safe}.

In Query~9, we learn a strategy \code{shielded\_efficient} under the shield using \stratego~\cite{DBLP:conf/tacas/DavidJLMT15}.
This strategy keeps the ball in the air without excessive hitting, as shown by the output of Query~10 in \cref{fig:shielded_efficient}.
The result of Query~11 shows the expected cost, and Query~12 shows that the safety property holds with high confidence: None of the $10\,000$ runs were unsafe.

\subsection{Further Examples}
\label{sect:benchmarks}

State-space transformations can be used to synthesize a shield more efficiently~\cite{DBLP:conf/vecos/BrorholtHLS24}.
Since \uppaal supports function calls, transformations can also be applied in our tool by modifying the model. We demonstrate that in \cref{appendix:transformations}.

Next, we show quantitative results of the shield synthesis and subsequent shield reduction, for which we also use three additional models.
Firstly, the \emph{boost converter}~\cite{DBLP:conf/vecos/BrorholtJLLS23} models a real circuit for stepping up the voltage of a direct current (DC) input.
The controller must keep the voltage close to a reference value, without exceeding safe bounds for the voltage and current.
The state space is continuous, with significant random variation in the outcome of actions.

In the \emph{random walk} model~\cite{DBLP:conf/vecos/BrorholtJLLS23, DBLP:conf/isola/Jaeger0BLJ20}, the player must travel a certain distance before time runs out by choosing between a fast but expensive and a slow but cheap action.
The state space is continuous and the outcomes of actions follow uniform distributions.

In the \emph{water tank} model inspired from~\cite{DBLP:conf/aaai/AlshiekhBEKNT18}, a tank must be kept from overflowing or running dry.
Water flows from the tank at a rate that varies periodically.
At each time step, the player can control the inflow by switching a pump on or off.
The state space is discrete.

We show results for computing and reducing shields in \cref{tab:evaluation}.
The \emph{water tank} is fully deterministic, and the \emph{bouncing ball} only has low-variance stochastic behavior. 
The \emph{boost converter} and \emph{random walk} have a high variance in action outcomes, which is why we use $m=20$ simulation runs per sampled state.
We evaluated the shields statistically and found no unsafe runs in $10\,000$ trials.
The reduction yields significantly smaller representations at acceptable run time.

\begin{table}[tb]
    \caption{Computation time and sizes for synthesizing and reducing shields for three models. The original size is the number of cells, whereas the reduced size is the number of regions. All shields were statistically evaluated to be at least 99.47\% safe with a confidence interval of 99\% (no unsafe runs observed).}
    \label{tab:evaluation}
    \centering
        \setlength\tabcolsep{1mm}
    \begin{tabular}{l c r r r r r}
        \toprule
        \cc{Model}      & \cc{$n$} & \cc{$m$}  & \cc{Synthesis time} & \cc{Size}   & \cc{Reduction time} & \cc{Reduced size} \\
        \midrule
        Bouncing ball   & 3        & 1         & 218s                & 1\,430\,000 & 53s                 & 2972              \\
        Boost converter & 3        & 20        & 1\,430s             & 136\,800    & 21s                 & 571               \\
        Random walk     & 4        & 20        & 82s                 & 40\,000     & 1.5s                & 60                \\
        Water tank      & 3        & 1         & 0.1s                & 168         & 0.1s                & 24                \\
        \bottomrule
    \end{tabular}
\end{table}

\section{Conclusion}

We have described our implementation of the shield synthesis algorithm from~\cite{DBLP:conf/vecos/BrorholtJLLS23} in the tool \ourtool.
Our tool can work with rich inputs modeled in \uppaal.
We have also presented the \caap algorithm to reduce the shield representation significantly, which is crucial for deployment on an embedded device.

We see several directions for future integration into \uppaal.
As discussed, our implementation does not apply \emph{systematic} sampling for random dynamics; however, we think that many sources of randomness in \uppaal models can be handled systematically.
Currently, the reduction algorithm \caap is implemented as a standalone tool, but it would be useful to also integrate it directly with \uppaal.
During development, we found it helpful to visualize shields, as in \cref{fig:bbshields}, which could be offered in the user interface.
In the same line, an explanation why a state is marked unsafe in a shield would help in debugging a model.

\begin{credits}
\subsubsection{\ackname} This research was partly supported by the Independent Research Fund Denmark under reference number 10.46540/3120-00041B, DIREC - Digital Research Centre Denmark under reference number 9142-0001B, and the Villum Investigator Grant S4OS under reference number 37819.

\subsubsection{\discintname}
The \uppaal tool is developed by Aalborg University.
\end{credits}

%
%
\bibliographystyle{splncs04}
\bibliography{bibliography}

\clearpage

\appendix

\section{Query Syntax}
\label{appendix:query_syntax}

\cref{eq:query} shows a query to compute the shield in \cref{fig:leave_bounds} for the \emph{bouncing ball}.
\begin{equation}\begin{aligned}
    &\overbrace{\texttt{ strategy \texttt{\color{belizehole} shield} =}}^{\text{Declaration}}
    ~\texttt{acontrol:}
    ~\overbrace{\texttt{A[] !{\color{belizehole} Ball}.{\color{belizehole} Stop}}}^{\text{Desired invariant}} \\
    &\qquad\underbrace{\texttt{{
        \{}
        {\color{belizehole} v}[-13, 13]:1300,
        {\color{belizehole} p}[0, 11]:550,
        {\color{belizehole} Ball}.location
        \}
    }}_{\text{Parameters of the grid}}
    \label{eq:query}
\end{aligned}\end{equation}

In \uppaal, strategies are first-class objects of the query language~\cite{DBLP:conf/tacas/DavidJLMT15}.
In this case, we declare a strategy named \code{shield}.
The query starts with the keyword \code{acontrol} to indicate synthesis of a control strategy by \emph{approximating} a transition system, as previously described.
Next comes the safety property, which must be an invariant property, indicated by a (mandatory) \code{A[]} prefix, followed by an expression in the \uppaal language.
Finally, the query expects a description of the grid.
The user must list all relevant variables of the system; all other variables are ignored, as described in \cref{sect:missing_variables}.
For each relevant variable, we expect lower and upper bounds of the state space as well as the number of cells.
In the example, the velocity~\code{v} is bounded to $\interval{-13}{13}$ and we ask for 1300 cells in that dimension, corresponding to a cell diameter of $0.02$.
The location of a component is a special case of a (discrete) variable and requires no further parameters.
In the example, we want to keep track of \code{Ball.location}, for which the tool automatically infers that there are two possible values.
This results in a grid of $1300 \times 550 \times 2 = 1\,430\,000$ cells.

\medskip

We note that, for the \emph{bouncing ball} model, a coarser grid yields a less permissive shield, or even a shield where no action is considered safe (i.e., $\safecells$ is empty).
From the plots, we see that neighboring cells often have the same allowed actions.
In fact, every cell where the ball is in location \uppLoc{Stop} (half of all cells) is immediately marked as unsafe.
In the next section, we describe a general method to obtaining a compact shield representation.

\section{Nonperiodic Controllers}
\label{appendix:nonperiodic}

The previous work~\cite{DBLP:conf/vecos/BrorholtJLLS23} focused on systems with periodic control, i.e. there is a constant amount of time between player actions.
In contrast, \ourtool allows the time between player actions to vary according to the environment.
As described in \cref{sect:reachability}, it does so by running the simulation until the player faces a new choice.
We demonstrate this by modifying the player component from \cref{fig:player}, which uses guards and invariants to ensure that the player can act exactly when \code{x==0.1}.

\begin{figure}[h]
    \begin{subfigure}[h]{0.49\textwidth}
        \centering
        \includegraphics[width=0.8\linewidth]{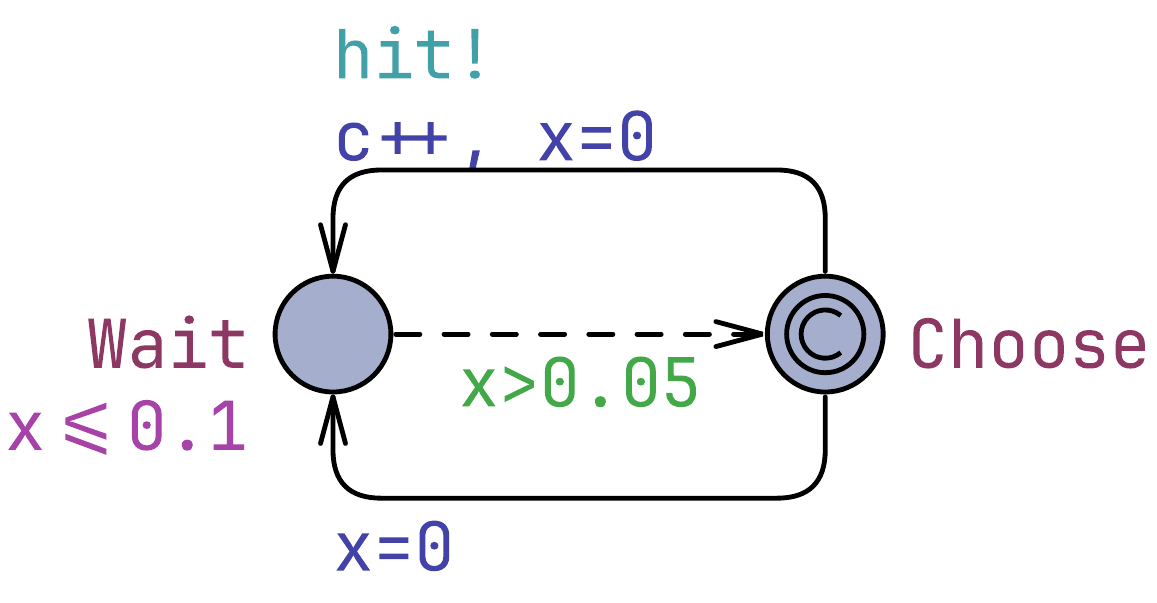}
        \caption{Stochastic player component. Compared to \cref{fig:player}, the guard has been updated to \uppGuard{x>0.05}. The ball component remains unchanged.}
        \label{fig:player_stochastic}
    \end{subfigure}
    \hfill
    \begin{subfigure}[h]{0.49\textwidth}
        \centering
        \includesvg[width=\linewidth]{Graphics/BB_stochastic_period.svg}
        \caption{Resulting shield.}
        \label{fig:rand_period_shield}
    \end{subfigure}
    \caption{Variant of the \emph{bouncing ball} with uniformly random decision periods.}
\end{figure}

The guard on the edge between \uppLoc{Wait} and \uppLoc{Choose} is changed to allow the system to stay in \uppLoc{Wait} for any period in the interval $\linterval{0.05}{0.1}$ (see \cref{fig:player_stochastic}).
For the modified system, we obtain a slightly different shield, as shown in \cref{fig:rand_period_shield}.

\section{\caap algorithm}
\label{appendix:caapalgorithm}

Let \( (p^{\min}, p^{\max}) \) define a region.
We then want to find a vector \( \Delta_p \in \mathbb{Z}^k \) such that \( (p^{\min}, p^{\min} + \Delta_p) \) defines a region that obeys the three expansion rules and is (locally) maximal, in the sense that increasing it in any dimension would violate at least one of the expansion rules.
Note that a vector \( \Delta_p = p^{\max} - p^{\min} \) satisfies the expansion rules trivially but is possibly not maximal.
Thus, a solution is guaranteed to exist.
However, note that there is not necessarily a unique maximal solution, and that the set of solutions is not convex, i.e., there may exist solutions~$\Delta_p^1$ and~$\Delta_p^2$ such that $\Delta_p^1 \le \Delta_p^2$ but no other~$\Delta_p'$ with $\Delta_p^1 \le \Delta_p' \le \Delta_p^2$ satisfies the expansion rules.
Formally:

\begin{definition}[Expansion vector \( \Delta _p \)]
\label{def:deltaP}
  Given \( p^{\min} \in \mathbb{Z}^k \), a decision tree \Tcal over a \( k \)-dimensional state space, and a set $\partit$ of fixed regions, $\Delta_{p} \in \mathbb{Z}^{k}$ is a vector such that for $p^{\max} = p^{\min} + \Delta_{p}$ the region $\region = (p^{\min}, p^{\max})$ does not violate any of the expansion rules in \cref{def:expansionRules} and for any vector \( \Delta'_p = (\Delta_{p_{1}}, \ldots, \Delta_{p_{i}} + 1, \ldots, \Delta_ {p_k}) \) at least one of the rules is violated.
\end {definition}

In \cref{alg:caap}, we formally present our greedy approach to finding $\Delta_{p}$.
It that starts with $\Delta_{p} = p^{\max} - p^{\min}$ for some region $\region = (p^{\min}, p^{\max})$.
It then iteratively selects a dimension~$d$ and attempts to increment the $d$-th entry of~$\Delta_{p}$.
For that, we define the candidate region $\region' = (p^{\min}, p^{\min} + \Delta_{p})$ and check the rules~\ref{it:rule1} and~\ref{it:rule2}.
If any of them is violated, we mark the corresponding dimension~$d$ as exhausted, roll back the increment, and continue with a new dimension not marked as exhausted yet, until none is left.

We implement the nondeterminism in line~\ref{line:choose_region} by prioritizing lower regions; this accounts for only expanding the upper bounds.
On the other hand, we implement the nondeterminism in line~\ref{line:choose_dimension} with a uniformly random choice.

\begin{algorithm}[t]
    \caption{\caap}\label{alg:caap}

    \begin{algorithmic}[1]
        \Require{%
            $\Tcal$: A binary decision tree inducing the partitioning $\partit_{\Tcal}$
        }
        \Ensure{%
            $\partit'$: A partitioning
        }
        \State{$\partit' \gets \{\}$}
        \State{Initialize matrix $M$ from $\partit_{\Tcal}$}

        \While{$\partit'$ does not cover $\partit _\Tcal$}

            \State{%
                $(p^{\min}, p^{\max}) \gets$ select an unexplored region from $M$ \label{line:choose_region}
            }

            \State{%
                $\Delta_p \gets p^{\max} - p^{\min}$
            }
            \State{%
                $\Delta'_p \gets \Delta_p$
            }

            \While{%
                not all dimensions have been exhausted
            }

                \State{$d \gets$ select a non-exhausted dimension}\label{line:choose_dimension}
                \State{$\Delta'_{p_{d}} \gets \Delta_{p_{d}} + 1$}
                \State{%
                    $\region' \gets (p^{\max}, p^{\max} + \Delta'_p)$
                }

                \If{%
                    $\region'$ violates rules~\ref{it:rule1} or~\ref{it:rule2}
                }
                    \State{%
                        $\Delta'_{p_{d}} \gets \Delta_{p_{d}}$
                    }
                    \State{mark $d$ as exhausted}

                \ElsIf{%
                    $\region'$ violates Rule 3 (\cref{def:expansionRules})
                }
                    \State{%
                        $\Delta''_p \gets \Call{Repair}{\Delta'_p, d}$
                    }
                    \If{repair was successful}
                        \State{$\Delta'_p \gets \Delta''_p$}
                    \Else%
                        \State{$\Delta'_p \gets \Delta_p$}
                        \State{mark $d$ as exhausted}
                    \EndIf%

                \Else
                    \State{%
                        \( \Delta_p \gets \Delta'_p \)
                    }
                \EndIf%

            \EndWhile%

            \State{%
                $\region \gets$ region defined by $(p^{\min}, p^{\min} + \Delta_p)$ according to $M$
            }
            \State{%
                $\partit' \gets \partit' \cup \{\region\}$
            }

        \EndWhile%

        \State{\textbf{return} $\partit'$}

    \end{algorithmic}
\end{algorithm}

As mentioned above, the set of solutions is not convex.
Correspondingly, if Rule~\ref{it:rule3} is violated, the algorithm initiates an attempt at \emph{repairing} the candidate expansion by continuing the expansion to the largest bound in the expansion dimension of any of the broken regions.
This way, we check whether the violation can be overcome by simply expanding more aggressively.
For conciseness, we do not describe the operation \textsc{Repair} further.
When all dimensions have been exhausted, \( \Delta _p \) adheres to \cref {def:deltaP}.

We note that the algorithm is not guaranteed to find a local optimum.
One reason is that the repair only expands in one dimension.
This choice is deliberate to keep the algorithm efficient and avoid a combinatorial explosion.

\section{State-Space Transformation}
\label{appendix:transformations}

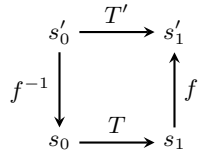
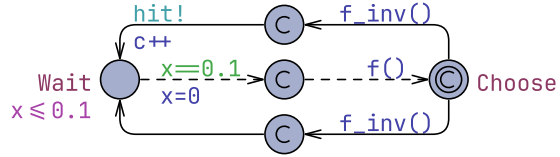
\begin{figure}[htb]
    \centering
    \begin{subfigure}[b]{0.31\linewidth}
        \centering
        \begin{tikzpicture}[t/.style={->,thick,>=stealth}]
	\node (s0') {$s_0'$};
	\node[right=of s0'] (s1') {$s_1'$};
	\node[below=of s0'] (s0) {$s_0$};
	\node[at=(s1'|-s0)] (s1) {$s_1$};
	\draw[t] (s0') to node[above] {$\transitionfunction'$} (s1');
	\draw[t] (s0') to node[left] {$f^{-1}$} (s0);
	\draw[t] (s0) to node[above] {$\transitionfunction$} (s1);
	\draw[t] (s1) to node[right] {$f$} (s1');
\end{tikzpicture}
 
        \caption{Commutative diagram.}
        \label{fig:commutative_diagram}
    \end{subfigure}
    \hfill
    \begin{subfigure}[b]{0.60\linewidth}
        \centering
        \includegraphics[width=\linewidth]{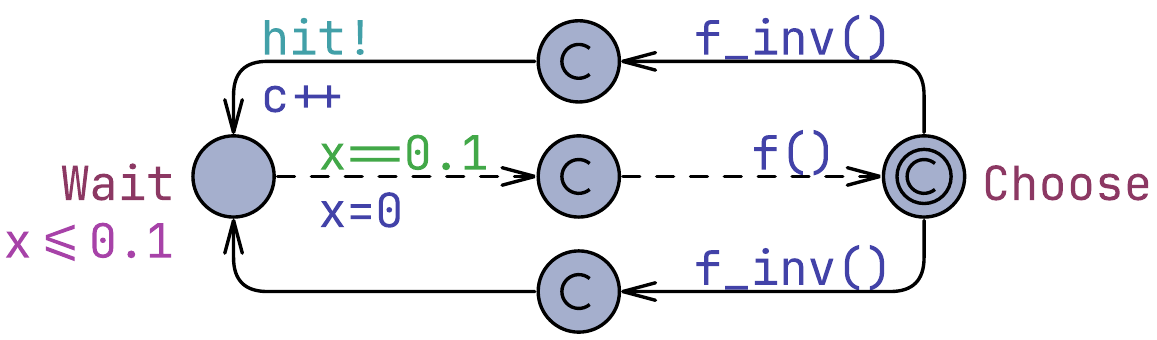}
        \caption{Augmented player component (cf.\ \cref{fig:player}).}
        \label{fig:trans_player}
    \end{subfigure}
    \caption{State-space transformations and template implementation in \uppaal.} \label{fig:trans}
\end{figure}

In~\cite{DBLP:conf/vecos/BrorholtHLS24}, we showed that state-space transformations can drastically reduce the synthesis time of a shield.
The idea is to define the grid in a transformed state space $\states' \subseteq \RR^{k'}$.
The method relies on a function $f \colon \states \to \states'$, mapping each state to a transformed state, and another function $f^{-1} \colon \states' \to \states$ mapping back.
\cref{fig:commutative_diagram} shows how to compute successors in~$\states'$, which is required for approximating reachability as described in \cref{sect:reachability}.
Given a state $s_0' \in \states'$, we wish to find a possible successor $s_1'$ for a given action.
The transition function~$\transitionfunction$ of the EMDP is defined over~$\states$.
Hence, we apply~$f^{-1}$ to obtain a corresponding state $s_0 \in \states$.
Then, we simulate~$\transitionfunction$ as before.
Finally, we apply~$f$ to obtain~$s_1'$.

The same method can be applied in \ourtool by modifying the model, which we show using the \emph{bouncing ball} model.
In~\cite{DBLP:conf/vecos/BrorholtHLS24}, the transformation uses the ball's mechanical energy~\code{e} instead of~\code{p}, with transformation function~$f(p, v) = (9.81 p + \frac{1}{2} v^2, v)$.

Thanks to the rich modeling language in \uppaal, it is easy to augment a \uppaal model to support state-space transformations.
We propose the following general modifications.
First, add corresponding new variables and implement the transformation functions as standard functions (e.g., called \code{f()} and \code{f\_inv()}) in the \uppaal language.
Then, modify the controller actions as illustrated in \cref{fig:trans_player}: add calls to \code{f\_inv()} immediately after the \uppLoc{Choose} location is left, and add calls to \code{f()} right before the \uppLoc{Choose} location is entered again.

\medskip

The following query ran for 5 seconds and produced a shield of just $2 \times 25 \times 26 =1300$ cells, which \caap can further reduce to just~$79$ regions.

\begin{equation*}\begin{aligned}
    &\texttt{strategy \texttt{\color{belizehole} safe} =}
    ~\texttt{acontrol:}
    ~\texttt{A[] !{\color{belizehole} Ball}.{\color{belizehole} Stop}} \\
    &\qquad\texttt{
        \{
        e[0, 100]:25,
        v[-13, 13]:26,
        {\color{belizehole} Ball}.location
        \}
    }
    \label{eq:trans_query}
\end{aligned}\end{equation*}

\section{From Regions to a Decision Tree}
\label{appendix:regionsToD}

The output of the \caap algorithm is a set of regions, each of which has an associated set of actions.
While this set of regions represents a shield, it is not efficient for querying at run time.
Hence, we aim to represent the set with another decision tree.
However, it is unlikely that the suggested partitioning can be perfectly represented by a decision tree.
For instance, the predicate in the root node always splits the whole state space, but we may have partially eliminated that split.
To that end, we propose a simple algorithm that constructs a new tree from a list of regions by recursively searching for a predicate that balances the task of splitting as few regions as possible while also dividing the regions into two nearly equal-sized subsets.
The resulting tree induces a partitioning that is finer than the partitioning used to create it.
Still, as we will see, the reduction gained from applying \caap to the original input is so significant that the cost of converting its output to a decision tree is negligible.

Since \caap does not guarantee optimal reduction but selects its expansion dimensions nondeterministically, we can achieve a better reduction by repeated application of the algorithm. That is, after obtaining a smaller
partitioning and converting it to a decision tree, we use that new tree as input to the algorithm once again.
This process is repeated until no significant reduction is observed.
Note that the process does not converge to a fixed-point due to the nondeterministic choices.
We found experimentally that the main reduction is achieved in the first application, and that the size of the output typically stabilizes after a few iterations.
Later repetitions are also less expensive due to the reduced input.

\end{document}